\documentclass[aps,prev,twocolumn,preprintnumbers,floatfix,nofootinbib]{revtex4-1}

\usepackage{graphicx}
\usepackage[ngerman,english]{babel}
\usepackage{bm}
\usepackage{times}
\usepackage{slashed}
\usepackage{mathtools}
\usepackage{color}
\usepackage{epsfig}
\usepackage{dcolumn}
\usepackage{textcomp}
\usepackage{color}
\newcommand\scalemath[2]{\scalebox{#1}{\mbox{\ensuremath{\displaystyle #2}}}}

\begin{document} 


\title{Discovering heavy U(1)-gauged Higgs bosons at the HL-LHC}

\author{Daniel A. Camargo$^{1,2}$}
\author{Michael Klasen$^{2}$}\email{michael.klasen@uni-muenster.de}
\author{Sybrand Zeinstra$^{2}$}

\affiliation{$^1$International Institute of Physics, Universidade Federal do Rio Grande do Norte, Campus Universitario, Lagoa Nova, Natal-RN 59078-970, Brazil}
\affiliation{$^2$Institut f\"ur Theoretische Physik, Westf\"alische Wilhelms-Universit\"at M\"unster, Wilhelm-Klemm-Stra\ss{}e 9, 48149 M\"unster, Germany}

\begin{abstract}

We determine the discovery potential of the Large Hadron Collider (LHC) at 13 TeV center-of-mass energy for a heavy scalar resonance in the dilepton channel. In particular, we consider the singlet-like heavy mass eigenstate of a mixed two Higgs doublet and scalar singlet model in the $U(1)_{B-L}$ extension of the Standard Model. We find that, despite the small coupling of the singlet scalar with the doublets, this heavy scalar can be discovered with 5$\sigma$ at the LHC with integrated luminosities of $\sim 300$ to $1400$ fb$^{-1}$ in the mass range between 500 GeV and 1 TeV.
\end{abstract}

\maketitle
\flushbottom

\section{Introduction}
\label{sec_introduction}

The discovery of a scalar resonance with a mass of 125 GeV at the Large Hadron Collider (LHC) has at first glance completed the Standard Model (SM) of particle physics \cite{Aad:2012tfa,Chatrchyan:2012xdj}.
However, it is well known that the abundance of dark matter in the Universe and the origin of neutrino masses cannot be explained within the SM. Both of these problems can, however, be solved by extending the Higgs sector and connecting it to dark matter and neutrinos through the so-called Higgs portal \cite{Patt:2006fw,Djouadi:2011aa}. Extensions containing two Higgs doublets and/or additional singlets have been shown to be particularly successful for neutrinos \cite{Antusch:2001vn,Atwood:2005bf,Liu:2016mpf,Cheung:2017lpv,Arcadi:2017wqi,Bertuzzo:2018ftf}, dark matter \cite{LopezHonorez:2006gr,Gustafsson:2007pc,Dolle:2009fn,Chao:2012pt,Goudelis:2013uca,Honorez:2010re,LopezHonorez:2010tb,Klasen:2013btp,Esch:2013rta,Klasen:2013ypa,Arhrib:2013ela,Esch:2014jpa,Bonilla:2014xba,Queiroz:2015utg,Arcadi:2018pfo}, with respect to the stringent limits on flavor-changing neutral currents (FCNCs) \cite{Ma:1998dx,Ma:2000cc,Ma:2002nn,Grimus:2009mm}, or all of the above \cite{Klasen:2013jpa,Klasen:2016vgl,Esch:2018ccs,Fiaschi:2018rky}. These extra scalars have been studied in the literature in the context of effective, simplified and UV-complete models. While for the first two the parameter space remains large and ambiguous, the parameters in UV-complete models are closely connected to each other, making their study more challenging, but also more consistent. 

$U(1)_X$ gauge extensions of the SM with an extended Higgs sector can explain the current relic abundance of dark matter in the Universe with scalar singlet, doublet or triplet dark matter \cite{Klasen:2016qux,Bauer:2018egk,Camargo:2019ukv} and the origin of neutrino masses through the seesaw mechanism \cite{Camargo:2018uzw}. Recently, LHC lower limits on the masses of the new gauge bosons predicted by different extra gauge symmetries have been imposed \cite{Jezo:2012rm,Jezo:2014wra,Camargo:2018klg}. Current dark matter experimental bounds on the $Z-Z'$ mixing angle have been shown to leave open some parameter space that can be explored by upcoming experiments \cite{Camargo:2019ukv}. Here, we turn to the scalar sector and explore the sensitivity of the LHC to heavy scalar resonances in the $U(1)_{B-L}$ gauge extension. Our study extends similar earlier studies for twin and composite Higgs models \cite{Buttazzo:2015bka} and the NMSSM \cite{Beskidt:2017dil}.

The LHC will remain the most powerful accelerator in the world for at least the next two decades. After the 2023-2025 upgrade, it will reach integrated luminosities of $\sim 1000$ fb$^{-1}$ or more \cite{Apollinari:2017cqg}. During the following high-luminosity (HL) phase, it will explore and potentially discover signals that were until then too small or hidden in large backgrounds.

\section{The Model}
\label{sec_model}

We consider a $U(1)_X$ extension of the SM gauge symmetry group that contains in addition to the SM particle content right-handed neutrinos, a two Higgs doublet model (2HDM), and a scalar singlet field. We focus on the
$U(1)_{B-L}$ symmetry because it is known to have many important implications in cosmology. In particular, the model is able to explain neutrino masses, while being at the same time free of FCNCs. It therefore resembles the type-I 2HDM with an additional singlet, where only one scalar doublet contributes to the SM fermion masses via the Yukawa Lagrangian
\begin{table*}[!t]
\caption{$U(1)_{B-L}$ charges for all fermions and scalars of our model. In particular, this assignment of charges is able to explain neutrino masses and the absence of flavor-changing currents in the type-I 2HDM.\vspace*{3mm}}
\label{cargas_u1_2hdm_tipoI}
\centering
\begin{tabular}{|c|ccccccccc|}
\hline 
Fields & $u_R$ & $d_R$ & $Q_L$ & $L_L$ & $e_R$ & $N_R$ & $\Phi _2$  & $\Phi_1$ & $\Phi_s$ \\ \hline
Charges & $u$ & $d$ & $\frac{(u+d)}{2}$ & $\frac{-3(u+d)}{2}$ & $-(2u+d)$ & $-(u+2d)$ & $\frac{(u-d)}{2}$ & $\frac{5u}{2} +\frac{7d}{2}$ & $2u+4d$ \\ \hline\hline
$U(1)_{B-L}$ & $1/3$ & $1/3$ & $1/3$ & $-1$ & $-1$ & $-1$ & $0$ & $2$ & 2 \\
\hline
\end{tabular}
\end{table*}
\begin{equation}
\begin{split}
\label{2hdm_tipoI_u1}
\mathcal{L} _{Y _{\text{2HDM}}} &= y_2 ^d \bar{Q} _L \Phi _2 d_R + y_2 ^u \bar{Q} _L \widetilde \Phi _2 u_R + y_2 ^e \bar{L} _L \Phi _2 e_R \\
&+ y^{D} \bar{L} _L \widetilde \Phi _2 N_R + Y^{M} \overline{(N_R)^{c}}\Phi_{s}N_R + h.c. \\
\end{split}
\end{equation}
Here, the scalar doublets are written as
\begin{equation}
\Phi _i = \begin{pmatrix} \phi ^+ _i \\ \left( v_i + \rho _i + i\eta _i \right)/ \sqrt{2}\end{pmatrix},
\end{equation}
while the scalar singlet is $\Phi_s = (v_s + \rho_s + i \eta_s)/\sqrt{2}$.

Models with additional $U(1)_X$ gauge symmetries and extended scalar sectors generally have reduced scalar potentials. Also our scalar potential includes fewer operators than the usual 2HDM with an additional singlet due to the $U(1)_{B-L}$ charge assignments of the scalar fields. In particular, the scalar doublets $\Phi_1$ and $\Phi_2$ have to transform differently under $U(1)_{B-L}$ in order to prevent FCNCs. At lower energies, this effect then plays the role of the $Z_2$ symmetry that is typically employed to stabilize dark matter. This setup also affects the mass spectrum of the scalar sector. The pure doublet and singlet-doublet parts of our scalar potential read
\begin{eqnarray}
 V \left( \Phi _1 , \Phi _2 \right) &=&
 m_{11} ^2 \Phi _1 ^\dagger \Phi _1 + m_{22} ^2 \Phi _2 ^\dagger \Phi _2 \nonumber\\
 &+& \frac{\lambda _1}{2} \left( \Phi _1 ^\dagger \Phi _1 \right) ^2 + \frac{\lambda _2}{2} \left( \Phi _2 ^\dagger \Phi _2 \right) ^2  \label{pot_2hdm_U1}\\
 &+& \lambda _3 \left( \Phi _1 ^\dagger \Phi _1 \right) \left( \Phi _2 ^\dagger \Phi _2 \right) + \lambda _4 \left( \Phi _1 ^\dagger \Phi _2 \right) \left( \Phi _2 ^\dagger \Phi _1 \right) , \nonumber \\
 V _s &=& m_s ^2 \Phi _s ^\dagger \Phi _s + \frac{\lambda _s}{2} \left( \Phi _s ^\dagger \Phi _s \right) ^2 + \mu _1 \Phi _1 ^\dagger \Phi _1 \Phi _s ^\dagger \Phi _s \nonumber \\
 &+&\mu _2 \Phi _2 ^\dagger \Phi _2 \Phi _s ^\dagger \Phi _s + \left( \mu_S \Phi _1 ^\dagger \Phi _2 \Phi _s + h.c. \right). \label{potscalarsinglet}
\end{eqnarray}
Note the absence of the $\lambda_5$ operator typical for 2HDMs,  which gives mass to the psedoscalar. This mass is generated here by the interaction of the scalar doublets with the singlet, requiring the parameters $\mu_i$ of the potential to be non-zero in order to avoid massless Goldstone bosons.

We are interested in a particular framework, in which there is a $m_H<m_h\ll m_S$ mass hierarchy. This framework will allow us to make consistent scans with a weakly coupled heavy scalar $S$ and light scalar $H$, while $h$ remains SM-like. In general, the neutral CP-even scalar mass eigenstates mix as
\begin{equation}
\begin{split}
\begin{pmatrix} h \\ H \\ S \end{pmatrix} = & \begin{pmatrix} c _{\alpha _2} & 0 & - s _{\alpha _2} \\ 0 & 1 & 0 \\ s _{\alpha _2} & 0 & c _{\alpha _2} \end{pmatrix} \begin{pmatrix} 1 & 0 & 0 \\ 0 & c _{\alpha _1} & - s _{\alpha _1} \\ 0 & s _{\alpha _1} & c _{\alpha _1} \end{pmatrix} \times \\
& \begin{pmatrix} c _\alpha & s _\alpha & 0 \\ - s _\alpha & c _\alpha & 0 \\ 0 & 0 & 1 \end{pmatrix} \begin{pmatrix} \rho _1 \\ \rho _2 \\ \rho _s \end{pmatrix} .
\end{split}
\end{equation}
The mixing angles $\alpha$, $\alpha _1$ and $\alpha _2$ depend on the parameters of the scalar potential and on the vacuum expectation values (VEVs) of the scalars. In the limit of $\mu_i \ll 1$ ($\alpha _1 , \alpha _2 \ll 1$), the angle $\alpha$ coincides with the mixing angle of the usual 2HDM with the $H$-$h$ mixing given by \cite{Campos:2017dgc}
\begin{eqnarray}
\left(\begin{array}{c}
H\\
h\\
\end{array}\right)\sim\left(\begin{array}{cc}
\cos\alpha & \sin\alpha\\
-\sin\alpha & \cos\alpha \\
\end{array}\right)\left(\begin{array}{c}
\phi_1\\
\phi_2\\
\end{array}\right)
\end{eqnarray}
and
\begin{equation}
\tan 2\alpha \sim \frac{2(\lambda_3+\lambda_4)v_1v_2}{\lambda_1 v_1^2 -\lambda_2 v_2^2}. 
\end{equation}
For the pseudoscalar and charged scalar we obtain
\begin{equation}
m_A^2 \sim \frac{\mu ( v_1 ^2 v_2 ^2 + v^2 v_s ^2 )}{\sqrt{2} v_1 v_2 v_s}
\label{pseudoscalar_mass}
\end{equation}
and
\begin{equation}
m_{H^+} ^2 \sim \frac{( \sqrt{2} \mu v_s - \lambda_4 v_1 v_2 ) v^2}{2 v_1 v_2}
\label{charged_scalar_mass}
\end{equation}
with $v^2 = v_1 ^2 + v_2 ^2 = (246\ \text{GeV}) ^2$. In the limit $\alpha_{1,2}\ll 1$, the scalar masses can be approximated by
\small\begin{eqnarray}
m_s^2 & \sim & \lambda_s v_s^2,\label{massescalars} \\
m_H^2  &\sim & \frac{1}{2}\left( \lambda_1 v_1^2 + \lambda_2 v_2^2 - \sqrt{(\lambda_1 v_1^2-\lambda_2 v_2^2)^2 +4(\lambda_3+\lambda_4)^2 v_1^2 v_2^2} \right), \nonumber\\ 
m_h^2  &\sim & \frac{1}{2}\left( \lambda_1 v_1^2 + \lambda_2 v_2^2 + \sqrt{(\lambda_1 v_1^2-\lambda_2 v_2^2)^2 +4(\lambda_3+\lambda_4)^2 v_1^2 v_2^2} \right),\nonumber
\end{eqnarray}\normalsize
where clearly the scalar singlet can be seen as if it had decoupled form the doublets. Its mass then depends only on its self-coupling $\lambda_s$ and VEV $v_s$.

\begin{figure}[h!]
\includegraphics[width=1\columnwidth]{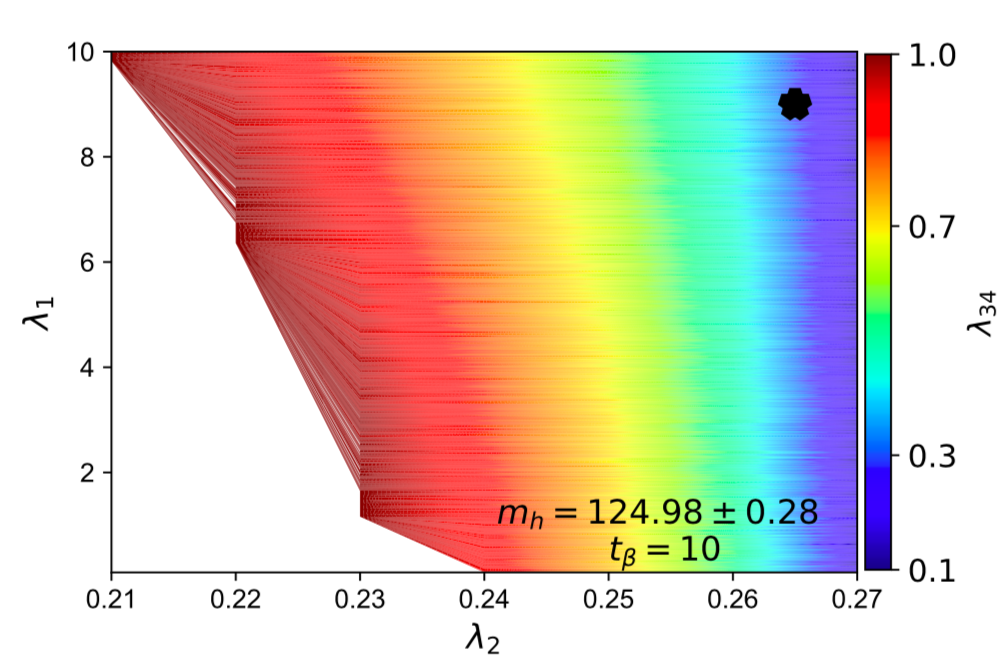}
\caption{Parameter space in the $\lambda_{1,2,34}$ plane compatible with current measurements of the SM Higgs boson mass $m_h$ for a ratio of Higgs doublet VEVs $t_\beta = \tan\beta=v_2/v_1=10$. Our benchmark point is shown in black.}
\label{fig:Hmass}
\end{figure}

As one can observe from Eqs.\ \eqref{massescalars}, the masses of the SM-like Higgs boson $h$ and the lighter boson $H$ are coupled, and their splitting depends in particular on the combined parameter $\lambda_{34}=\lambda_3+\lambda_4$. In Fig.\ \ref{fig:Hmass} we show the available parameter space in the $\lambda_{1,2,34}$ plane compatible with the current measurement of the SM-like Higgs boson mass $m_h$ for a ratio of the Higgs doublet VEVs $t_\beta = \tan\beta = v_2/v_1 = 10$. We stress that all scanned points shown here are compatible with the stability and perturbativity of the potential as required \cite{Xu:2017vpq,Chen:2018uim}. The black star denotes the benchmark point that we will consider in our phenomenological studies. The first conclusion that we can draw form Fig.\ \ref{fig:Hmass} is that the available parameter space is strongly affected by imposing the mass of the SM-like Higgs boson, requiring the parameter $\lambda_2$ to lie in the narrow window $[0.21-0.27]$ for $0.01<\lambda_1 < 10$ and $0.1<\lambda_{34} < 1$.

\begin{figure}
\includegraphics[width=1\columnwidth]{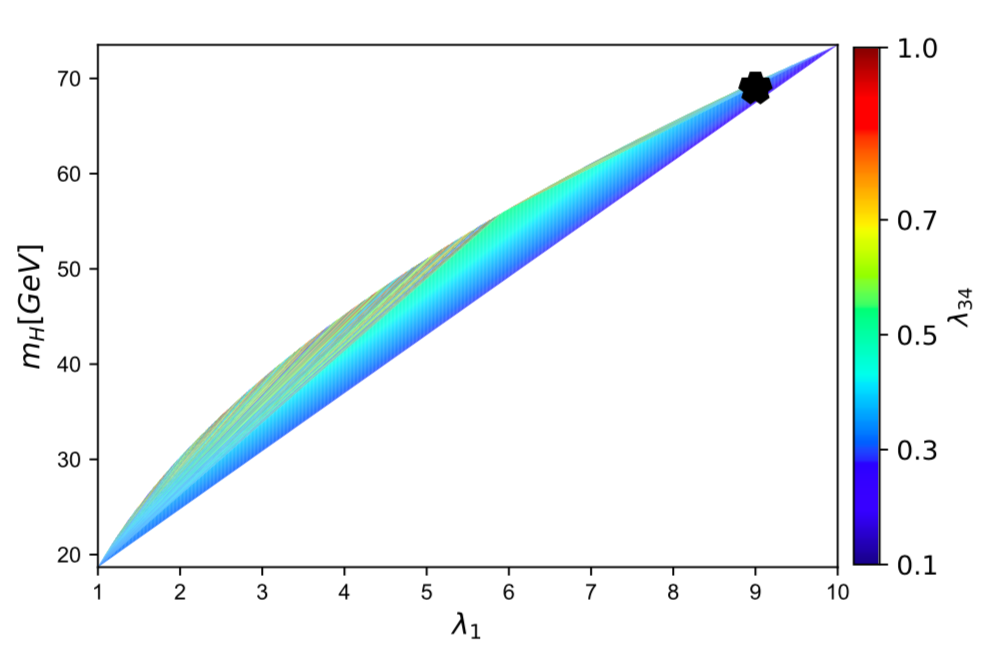}
\caption{Dependence of the mass of the lighter Higgs $H$ on the self-couplings $\lambda_1$ and $\lambda_{34}$. Our benchmark point is shown in black.}
\label{fig:H2mass}
\end{figure}

In Fig.\ \ref{fig:H2mass} we show the dependence of the mass of the lighter Higgs boson $H$ coming from the second doublet in the region compatible with the measured mass of the SM-like Higgs boson $h$. As one can see, $m_H$ depends almost linearly on $\lambda_1$ for small values of $\lambda_{34}<\lambda_1$ and fixed $m_h$. The black star denotes again our benchmark scenario, which we have chosen to avoid low-energy constraints coming from the $STU$ parameters. This is achieved by selecting a mass of the pseudoscalar Higgs boson $A$ in Eq. \eqref{pseudoscalar_mass} that is nearly degenerate with $m_H$. For our benchmark point with $t_\beta=10$, $v_s=5\times10^4$ GeV and $\mu=7\times 10^{-6}$, we obtain $m_A = 72$ GeV, which is indeed nearly degenerate with $m_H$ and thus compatible with the $STU$ requirements.

\begin{figure}
\includegraphics[width=1\columnwidth]{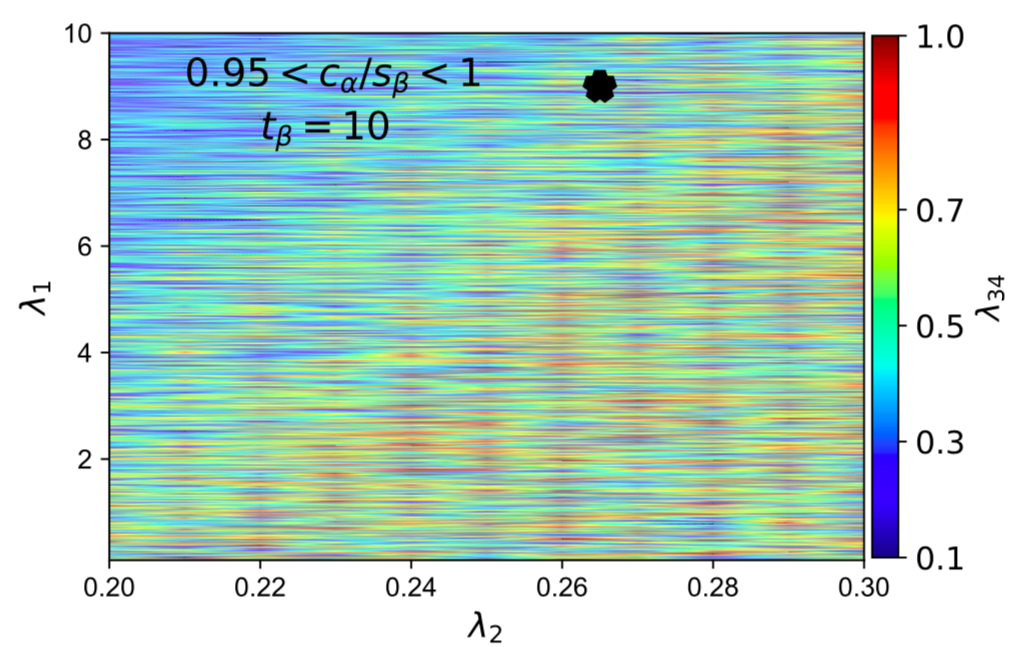}
\caption{Compatibility of the 2HDM with an additional scalar singlet with SM-like fermionic branching ratios of the Higgs boson $h$ for $t_\beta=10$. Our benchmark point is shown in black.}
\label{fig:alpha}
\end{figure}

The couplings of the SM fermions to the CP-even scalars are given in Tab.\ \ref{tableHiggs}. In the weak coupling limit of the heavy scalar singlet, they reduce to those given in Tab.\ \ref{tableHiggs2}.
\begin{table}
\caption{Scalar coupling constants of the SM fermions.\vspace*{3mm}}
\label{tableHiggs}
\centering
\begin{tabular}{|c|c|}
\hline
Vertex & Coupling constant\\
\hline\hline
$H\, t\bar{t},H\, b \bar{b} , H\, \tau \bar{\tau}$ & $\frac{\sin \alpha \cos \alpha _2 - \cos \alpha \sin \alpha _1 \sin \alpha _2}{\sin\beta}$ \\
\hline
$h\, t\bar{t},h\, b \bar{b} , h\, \tau \bar{\tau}$ & $\frac{\cos \alpha \cos \alpha _2 - \sin \alpha \sin \alpha _1 \sin \alpha _2}{\sin\beta}$ \\
\hline
$S\, t\bar{t},S\, b \bar{b} , S\, \tau \bar{\tau}$ & $\frac{\cos \alpha _1 \sin \alpha _2}{\sin\beta}$ \\
\hline
\end{tabular}
\end{table}
\begin{table}[!h]
\caption{Same as Tab.\ \ref{tableHiggs} in the limit $\alpha _1, \alpha _2 \ll 1$.\vspace*{3mm}}
\label{tableHiggs2}
\centering
\begin{tabular}{|c|c|}
\hline
Vertex & Coupling constant\\
\hline\hline
$H\, t\bar{t},H\, b \bar{b} , H\, \tau \bar{\tau}$ & $\frac{\sin \alpha}{\sin\beta}$ \\
\hline
$h\, t\bar{t},h\, b \bar{b} , h\, \tau \bar{\tau}$ & $\frac{\cos \alpha}{\sin\beta}$ \\
\hline
$S\, t\bar{t},S\, b \bar{b} , S\, \tau \bar{\tau}$ & $\frac{\sin \alpha _2}{\sin\beta}$ \\
\hline
\end{tabular}
\end{table}
As Fig.~\ref{fig:alpha} shows, SM-like branching ratios $0.95<\cos\alpha/\sin\beta<1$ of the Higgs boson $h$ can be achieved in all regions of our coupling parameter space.
We have also verified that for our benchmark point, shown again in black, the ligher CP-even Higgs boson $H$ has a branching ratio BR$(H\rightarrow b\bar{b})=97\%$, which makes its discovery difficult due to the large QCD backgrounds. 

Extensions of the gauge group can in principle have a large impact on the branching ratios of the scalars to gauge bosons, which scale with the gauge coupling constant. While this is not relevant for the light Higgs boson $H$, as $m_H<2m_{W,Z}$, it could well be of importance for the prime object of our interest, the heavy scalar $S$. In $U(1)_X$ extensions of the SM, kinetic mixing between the neutral gauge bosons occurs due the mixing of the field strength tensors $B _{\mu \nu}$ and $X _{\mu \nu}$ of $U(1)_{Y}$ and $U(1)_{B-L}$,
\begin{equation}
\mathcal{L} _{\rm gauge} =  - \frac{1}{4} B _{\mu \nu} B^{\mu \nu} + \frac{\epsilon}{2\, \cos \theta_W} X _{\mu \nu} B^{\mu \nu} - \frac{1}{4} X _{\mu \nu} X ^{\mu \nu}.
\label{Lgaugemix1}
\end{equation}
We set the kinetic mixing parameter to $\epsilon = 10^{-4}$ in accordance with the experimental constraints \cite{Hook:2010tw, Mambrini:2011dw} with the result that the branching ratios of the heavy scalar $S$ to SM gauge bosons are also suppressed, while the decay into two $Z'$ bosons is in principle allowed.

\begin{figure}
\includegraphics[width=1\columnwidth]{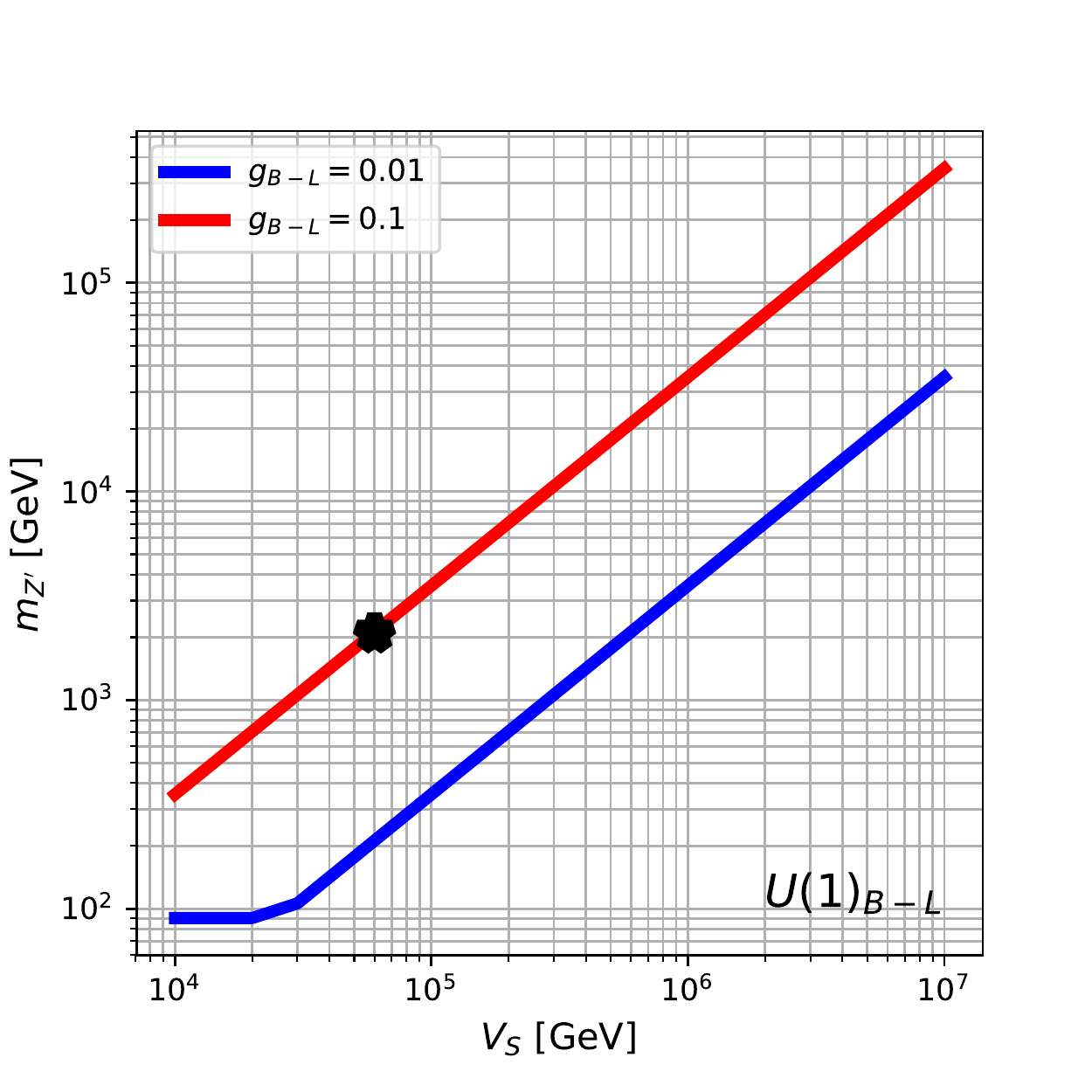}
\caption{Mass of the extra gauge boson $m_{Z'}$ as a function of the singlet VEV $v_s$ for two values of $g_{B-L} = 0.1$ and $g_{B-L} = 0.01$. Our benchmark point is shown in black.}
\label{fig:MZP}
\end{figure}

Additional mixing arises from the neutral gauge boson mass terms, since the scalar doublet $\Phi_1$ is charged under $B-L$ and contributes to the $Z^\prime$ mass. This mixing is, however, small in our case, since $v_s\gg v_1$. For large $v_s$, the new gauge boson mass is given by $m_{Z^\prime} \sim \frac{1}{2}g_{B-L}q_S v_s$, where $q_S=2$ is the charge of the singlet scalar under the new gauge symmetry. Its dependence on $v_s$ is shown in Fig.\ \ref{fig:MZP} for different values of the new gauge goupling $g_{B-L}$. We choose $g_{B-L}=0.1$ in accordance with the latest LHC limits  \cite{Camargo:2018rfi}.

In the following, we focus on the production and decay of the new heavy scalar $S$. Once the scalar bosons have mixed, they share all the production and decay channels allowed by the symmetries. The decay channels depend not only on the available phase space, but are also strongly model-dependent. Depending on the values of $v_s$ and $g_{B-L}$, scalar decays into $Z^\prime Z^\prime$ or $Z^\prime Z$ will be allowed by phase space. We have adopted scenarios, where those channels are closed by phase space, and focus instead on direct decays to SM particles. While these channels can provide information on the new gauge coupling, we leave their exploration for future work.

\begin{figure}
\includegraphics[width=1\columnwidth]{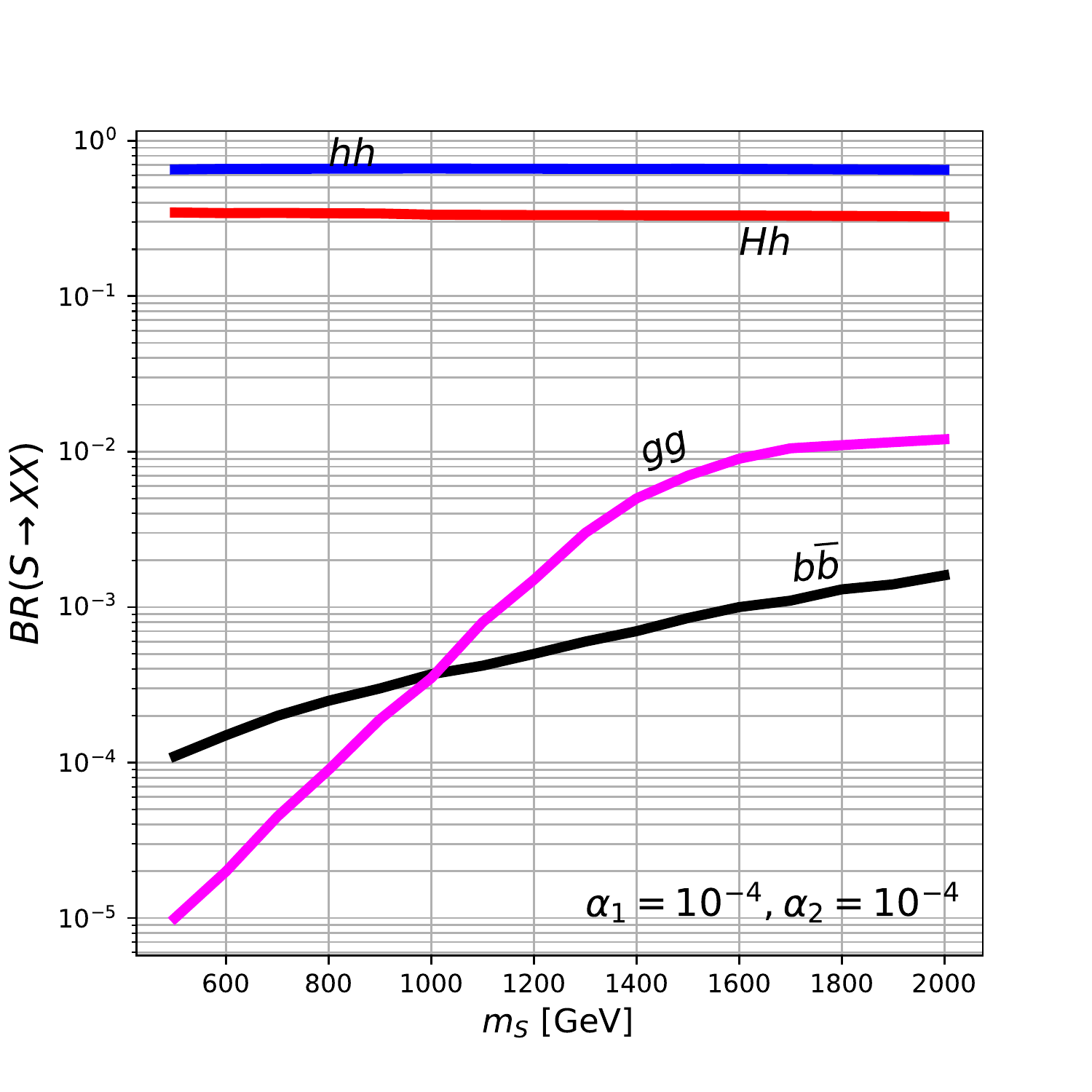}
\caption{Dominant branching ratios of the heavy scalar $S$ as a function of its mass for mixing angles $\alpha_{1}=\alpha_{2}=10^{-4}$.}
\label{fig:BR1}
\end{figure}
When the coupling of the heavy singlet $S$ to both doublets is very small ($\mu_i \sim 10^{-6}$ and $\alpha_{1,2} \sim 10^{-4}$), it decays primparily into $hh$ and $Hh$, as we can see in Fig.\ \ref{fig:BR1}. The reason is that the corresponding partial widths are proportional to $v_s$ (see Appendix) and remain competitive, despite the strong $\mu_i$ suppression, against the fermionic decay channels, that are proportional to $\sin \alpha_2$ (cf.\ Tab.\ \ref{tableHiggs2}).
\begin{figure}
\includegraphics[width=1\columnwidth]{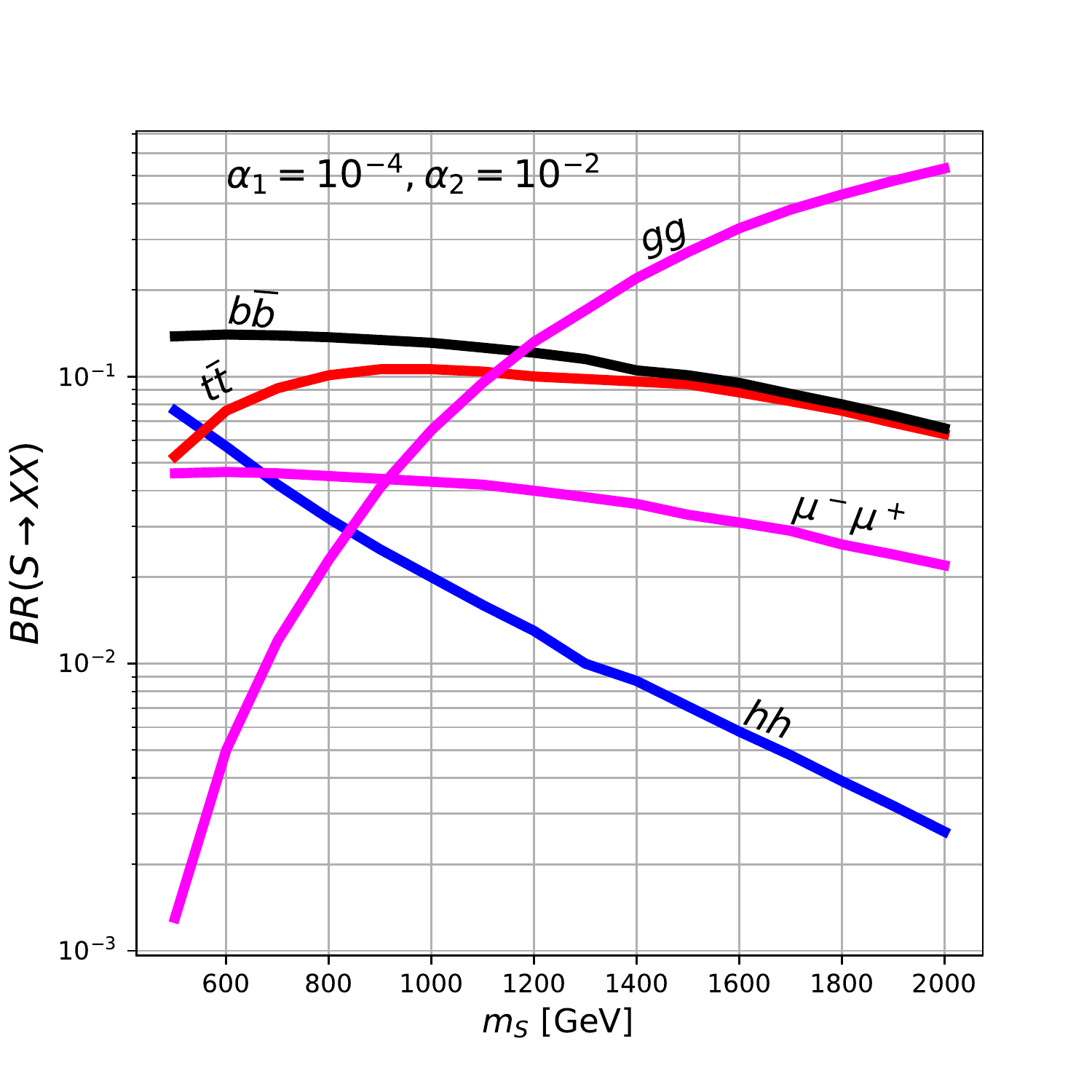}
\caption{Same as Fig.\ \ref{fig:BR1} for mixing angles $\alpha_{1}=10^{-4}$ and $\alpha_{2}=10^{-2}$.}
\label{fig:BR2}
\end{figure}
For larger couplings of the singlet to the second doublet ($\alpha_2\sim 10^{-2}$), the situation changes and the decay to dimuons becomes visible. This is the channel that we will exploit in the following.

\section{LHC phenomenology of a heavy singlet-like Higgs boson}

\begin{figure*}
\includegraphics[width=1\columnwidth]{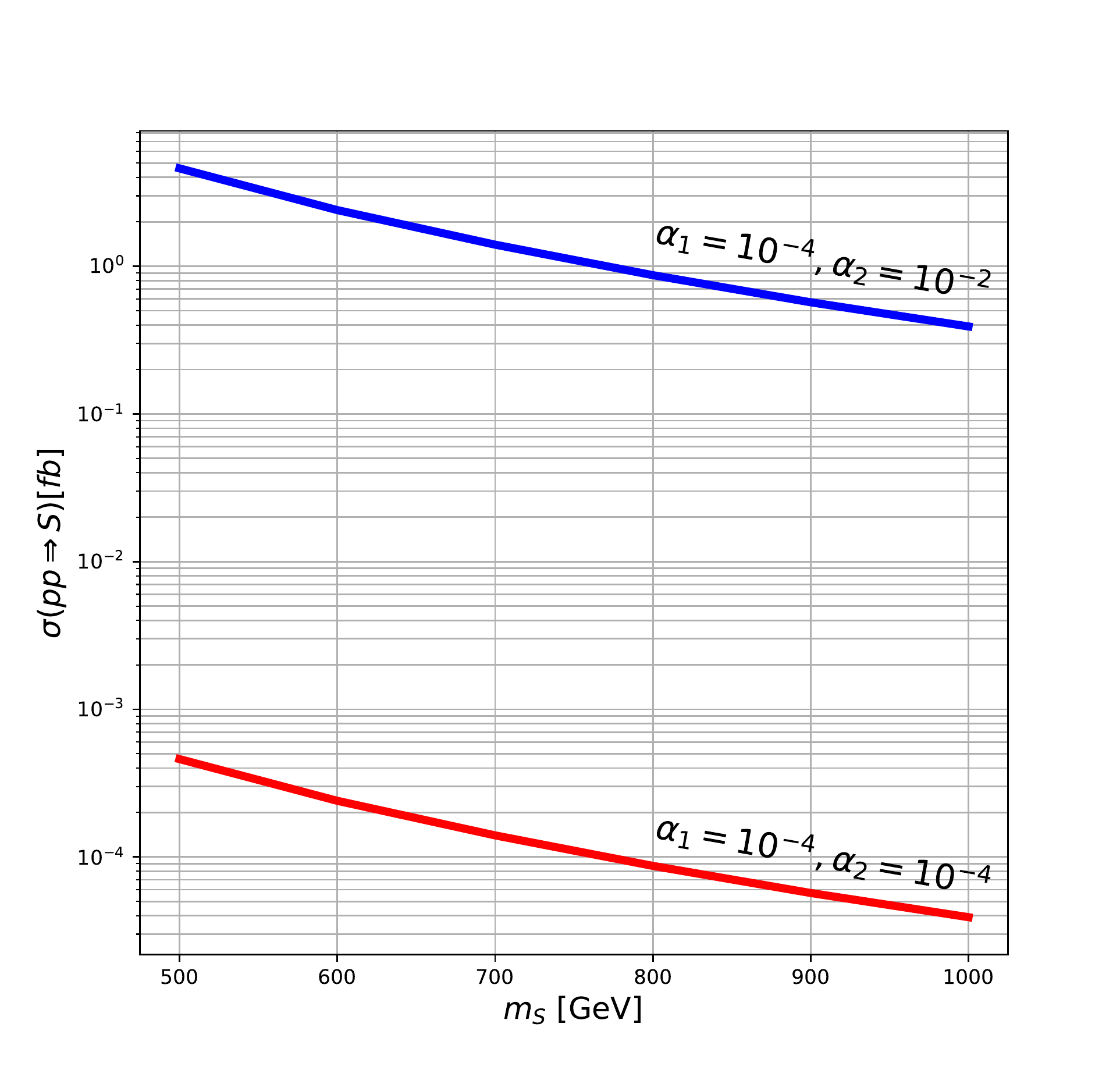}
\includegraphics[width=1\columnwidth]{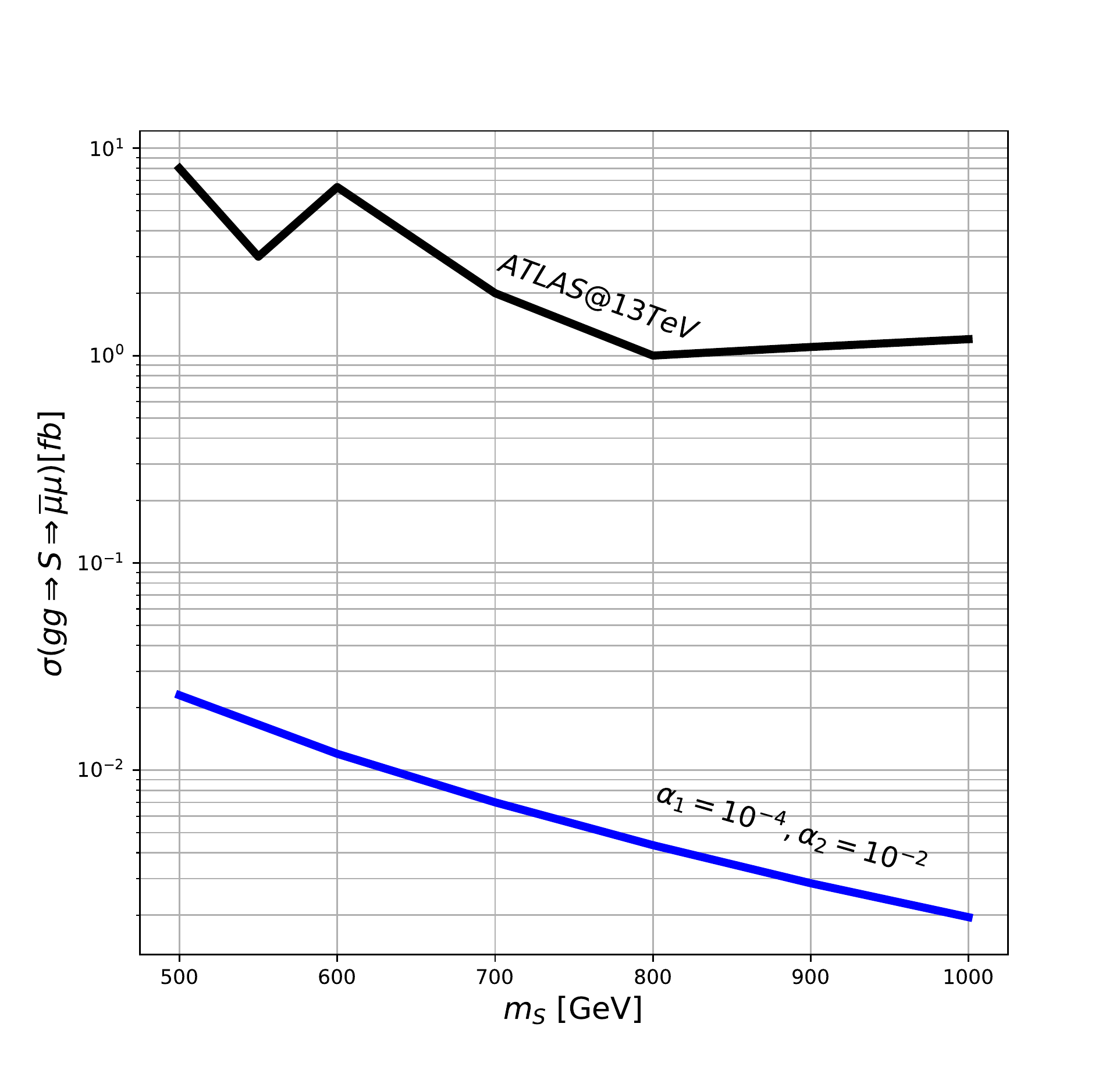}
\caption{Left: Total production cross sections of heavy scalars $S$ for scenarios with $\alpha_{1}=\alpha_{2}=10^{-4}$ and $\alpha_{1}=10^{-4}$, $\alpha_{2}=10^{-2}$. Right: Upper limit on the gluon-gluon fusion production of scalars decaying into dimuons imposed by the ATLAS collaboration (black) and our model prediction assuming $\alpha_{1}=10^{-4}$ and $\alpha_{2}=10^{-2}$ (blue).}
\label{fig:xsex}
\end{figure*}

Let us now explore the hypothesis of a heavy Higgs boson with mass at the TeV scale. We want to estimate its discovery prospects at the HL-LHC with 13 TeV center-of-mass energy. As we have adopted a scenario of a weakly coupled singlet-like scalar $S$, the mixing angles $\alpha_{1,2}$ have to be small. We first examine further its branching ratios in the two scenarios $\alpha_{1}=\alpha_{2}=10^{-4}$ and $\alpha_{1}=10^{-4}$, $\alpha_{2}=10^{-2}$. The dominant channels for both cases were shown in Figs.\ \ref{fig:BR1} and \ref{fig:BR2}. For the first case, the large branching ratios to $hh$ and $Hh$ seem promising at first sight. Unfortunately, the tiny production cross section makes this scenario inconceivable for the current and prospective luminosities of the LHC and HL-LHC, as one can see from Fig.\ \ref{fig:xsex} (left). Therefore, we focus on the second scenario in order to see if it is possible with the upgrade of the LHC and HL-LHC to explore the TeV-scale parameter space in the scalar sector of our model. For this scenario, where $\alpha_{1}=10^{-4}$ and $\alpha_{2}=10^{-2}$, we obtain a large dijet signal, which suffers, however, from a huge QCD background. The dimuon channel is therefore most promising, as its dominant background is SM $Z$-boson production.

To show the potential of the HL-LHC to discover a heavy Higgs boson that couples to a new gauge sector, we simulate the gluon fusion process for our benchmark model,
\begin{equation}
pp \rightarrow S \rightarrow l^- l^+,
\end{equation}
where $l^-$ and $l^+$ represent electrons or muons. We implement the $U(1)_{B-L}$ model and its interactions with the help of FeynRules \cite{Alloul:2013bka} and simulate the partonic events with MadGraph5 \cite{Alwall:2011uj}. One extra jet is also taken into account in the simulation in order to better estimate the kinematic distributions and cross sections. Hadronization and detector effects were taken
into account with the Pythia8 \cite{Sjostrand:2007gs} and Delphes3 \cite{deFavereau:2013fsa} interfaces to MadGraph5, respectively, within the kT-MLM jet matching scheme \cite{Mangano:2006rw}. The one-loop $S$ production through gluon-fusion has been implemented following the lines of Ref.\ \cite{Plehn:2009nd}. The relevant backgrounds for our signals are the $Z$, $h$, $\gamma$, $bZ$, $WZ$ and $b\bar{b}Z$ production processes. They were simulated with the same tools as those used in the signal simulation. In order to suppress these backgrounds and select the candidate signal events we adopt the following basic cuts:

\begin{eqnarray}
p_{T1,2}>20 \ {\rm GeV}&,& |\eta_\ell| < 2.5\\
\not{\!\!E}_T < 40\ {\rm GeV} &,& m_{2\ell} > 250 \ {\rm GeV},
\end{eqnarray}
where $p_{Tn}$ denotes the $n$-th hardest lepton of the event. We reject events with missing transverse energy larger than 40 GeV in order to eliminate $WZ$ and $t\bar{t}$ events. Backgrounds with bottom jets are efficiently cleaned up with lepton isolation criteria. Finally, a hard cut on the two-lepton invariant mass $m_S -40\;\hbox{GeV} < m_{\ell\ell} < m_S+40\;\hbox{GeV}$ helps to identify typical leptons from a heavy resonance decay. 

\begin{figure}[h!]
\includegraphics[width=1\columnwidth]{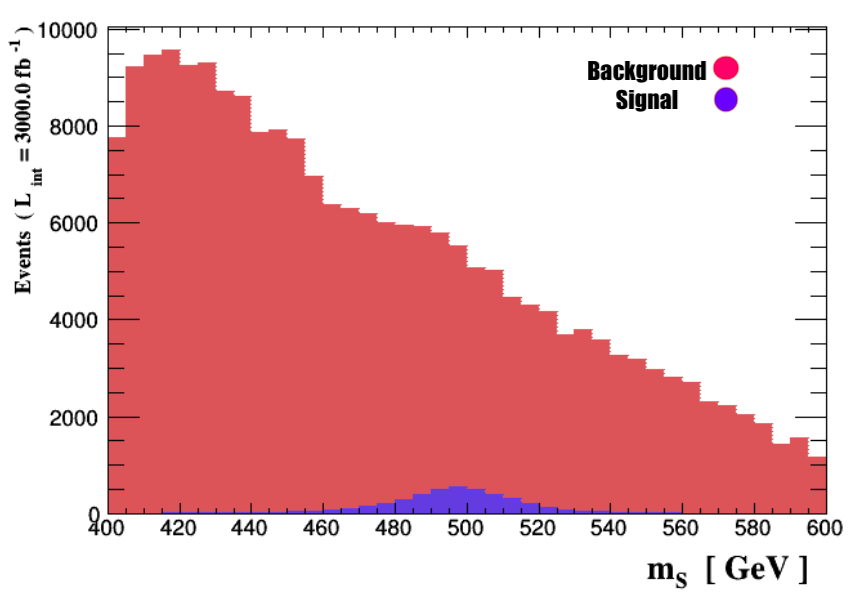}
\caption{Dilepton invariant mass distribution for a singlet-like scalar boson of mass $m_s=500$ GeV (blue) and its SM background (red) before cuts and for an integrated luminosity of 3000 fb$^{-1}$.}
\label{fig:reson}
\end{figure}

\begin{figure}[t!]
\includegraphics[width=1\columnwidth]{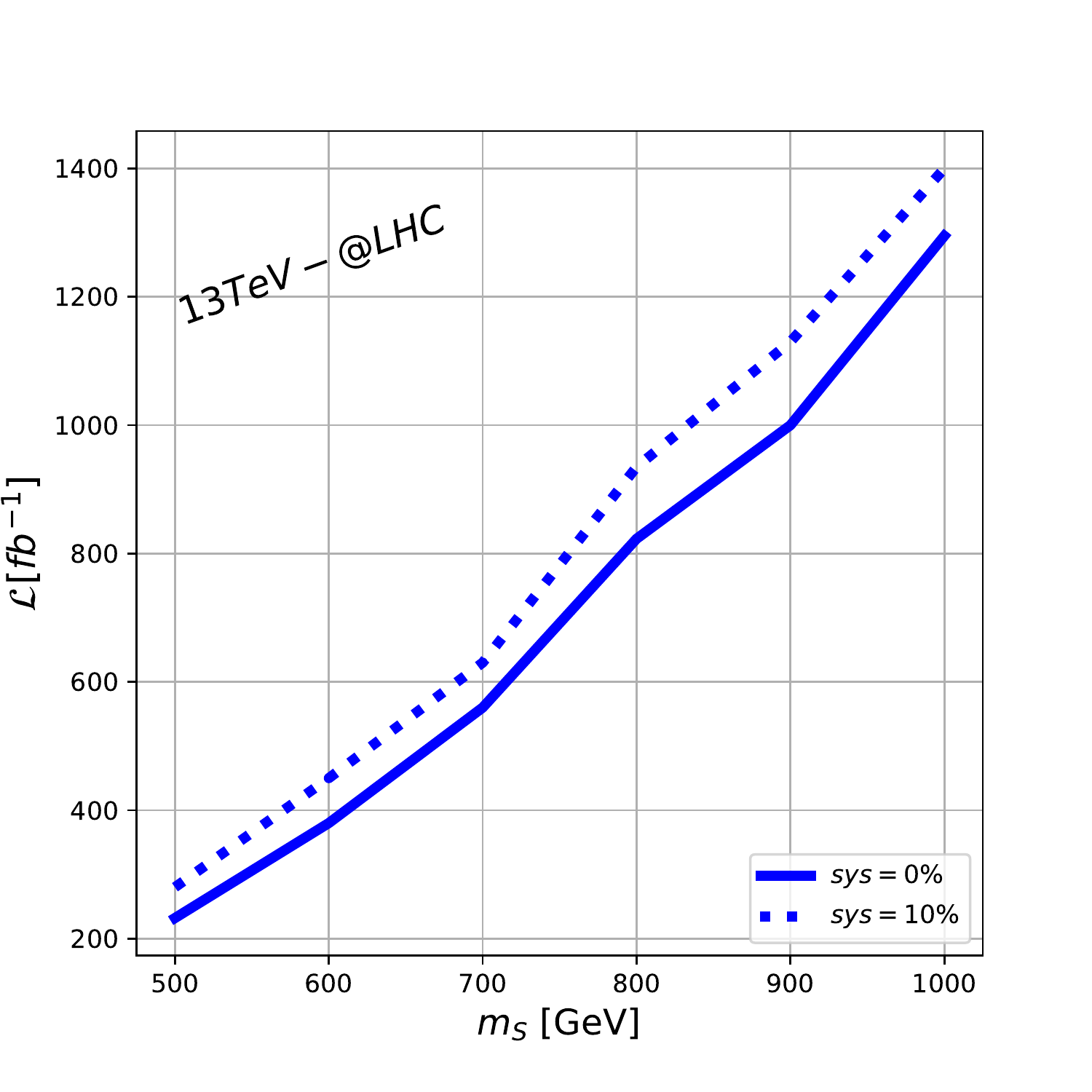}
\caption{Required luminosity of the HL-LHC at 13 TeV center-of-mass energy to discover a singlet-like Higgs boson $S$ with 5$\sigma$ as a function of its mass for an assumed systematic error or $10\%$ (dotted line) and $0\%$ (continuous line).}
\label{fig:lumi}
\end{figure}

In Fig.\ \ref{fig:reson}, we compare the signal of the heavy resonance of mass 500 GeV in the invariant mass distribution of the dilepton system to the background, which is still large before the kinematic cuts. After these cuts, the signal can be made visible at the level of 5$\sigma$ at the HL-LHC. Fig.\ \ref{fig:lumi} then shows the required luminosity at 13 TeV center-of-mass energy for this significance, computed with the significance metric $N_S/\sqrt{N_S+N_B+(xN_B)^2}$, where $x$ presents the assumed systematic error. The continuous line shows the ideal case with negligible systematic errors, while the dashed line assumes a systematic error of $10\%$. We observe that the current LHC with 100 fb$^{-1}$ of integrated luminosity cannot discover these singlet-like scalars, as their TeV-scale signals are still covered by too large backgrounds. However, in the near future, after the LHC luminosity upgrade scheduled for 2023 to 2025, it will become possible to explore singlet-like scalars up to masses of  $\sim 1$ TeV despite the fact that they are heavy,
only weakly coupled, constrained by the SM-like Higgs mass and branching ratios and are derived from UV-complete models with additional theoretical constraints.

\section{Conclusions}
\label{sec_model}
We have discussed the phenomenology of a TeV-scale scalar field living in a $U(1)_{B-L}$ gauge extension of the SM that contains both a 2HDM and a scalar singlet. The model is known to explain neutrino masses in a flavor-safe framework and can be extended in a straightforward way to also explain dark matter, {\em e.g.} by the addition of a vector-like fermion. In this case the heavy scalar would be the witness of an underlying broken gauge symmetry rather than an ad-hoc $Z_2$ symmetry typically invoked to stabilize dark matter. We focused on the scenario in which the singlet scalar is weakly coupled to the 2HDM sector. In this scenario, the known mass and branching ratios of the SM-like Higgs boson can be used to restrict several of the model parameters. We studied the discovery reach of the LHC with a center-of-mass energy of 13 TeV for such a heavy singlet-like scalar decaying to dileptons. We found that the HL-LHC with integrated luminosities of $\sim 1400$ fb$^{-1}$ allows for a discovery of these heavy Higgs bosons up to masses of about 1 TeV.    

\acknowledgments
D.C.\ thanks the University of M\"unster for hospitality and financial support through a WWU fellowship as well as MEC and UFRN. The work of M.K.\ and S.Z.\ is supported by the DFG through the Research Training Network 2149 ``Strong and weak interactions - from hadrons to dark matter''.

\section{Appendix}
\label{appendix}
\paragraph*{Partial widths of singlet-like heavy Higgs bosons}

The partial widths of the heavy scalar $S$ for the decay channels relevant to our study ($\mu\bar{\mu},\ b\bar{b},\, hH,\, ZZ,\, Z^{\prime}Z^{\prime}$ and $Z^{\prime}Z$) in the small coupling limit ($\alpha_{1,2} \ll 1$) are given by:

\small\begin{equation}
\begin{gathered}
\Gamma_{S\mu\bar{\mu}}=\frac{\sqrt{m_S^2(m_S^2-4m_\mu^2)}3s_\alpha^2(4m_\mu^2-m_S^2)}{8\pi s_\beta^2 m_S^3}
\end{gathered}
\end{equation}

\small\begin{equation}
\begin{gathered}
\Gamma_{Sb\bar{b}}=\frac{\sqrt{m_S^2(m_S^2-4m_b^2)}3s_\alpha^2(4m_b^2-m_S^2)}{8\pi s_\beta^2 m_S^3}
\end{gathered}
\end{equation}

\small\begin{equation}
\begin{gathered}
\Gamma_{ShH}=\frac{\sqrt{m_h^4-2m_h^2m_H^2+m_H^4-2m_h^2m_S^2-2m_H^2m_S^2+m_S^4}}{64\pi m_S^3}\times\\
(\mu c_\alpha^2 +\mu_S s_\alpha^2 - \mu_{1}s_\alpha v_s - \mu_{2}s_\alpha v_s)^2
\end{gathered}
\end{equation}

\small\begin{equation}
\begin{gathered}
\Gamma_{SZZ}=g_{B-L}^4\frac{\sqrt{m_S^2(m_S^2-4m_Z^2)} (m_S^4-4m_S^2m_Z^2+12m_Z^2)s_\xi^4v_s^2}{128 m_Z^4\pi m_S^3}
\end{gathered}
\end{equation}

\small\begin{equation}
\begin{gathered}
\Gamma_{SZ^{\prime}Z^{\prime}}=g_{B-L}^4\frac{\sqrt{m_S^2(m_S^2-4m_Z^{\prime 2})} (m_S^4-4m_S^2m_Z^2+12m_Z^{\prime 2})c_\xi^4v_s^2}{128 m_Z^{\prime 4}\pi m_S^3}
\end{gathered}
\end{equation}

\small\begin{equation}
\begin{gathered}
\Gamma_{SZ^{\prime}Z}=g_{B-L}^4\frac{\sqrt{m_S^4 - 2m_S^2 m_Z^2 +m_Z^4 -2m_S^2 m_Z^{\prime 2} - 2m_Z^2m_Z^{\prime 2} + m_Z^{\prime 4}}}{256 m_Z^2 m_Z^{\prime 2}\pi m_S^3}\\
\times (m_S^4-2m_S^2m_Z^2+12m_Z^{\prime 4} - 2m_S^2m_Z^{\prime 2}+10m_Z^2m_Z^{\prime 2} + m_Z^{\prime 4})q_S^4s_\xi^2c_\xi^2v_s^2
\end{gathered}
\end{equation}
\normalsize
Here, we have defined $s_\xi = \sin\xi$ and $c_\xi = \cos\xi$, $s_\beta = \sin\beta$ and $s_\alpha = \sin\alpha$ for simplicity. 

\paragraph*{Gauge kinetic terms and gauge boson masses}

In the canonical basis, the gauge covariant derivative for small $\epsilon$ reads
\begin{equation}
\label{der_cov_u1_diag}
\small
D_\mu = \partial _\mu + ig T^a W_\mu ^a + ig ' \frac{Q_Y}{2} B _{\mu} + \frac{i}{2} \left( g ' \frac{\epsilon Q_{Y}}{\cos \theta _W} + g_X Q_X \right) X_\mu
\end{equation}
or explicitly
\begin{equation}
\small
 D_\mu = \scalemath{0.8}{\partial _\mu + \frac{i}{2} \begin{pmatrix} g W_\mu ^3 + g ' Q_{Y} B_\mu + G_X X_\mu & g \sqrt{2} W_\mu ^+ \\ g \sqrt{2} W_\mu ^- & - g W_\mu ^3 + g ' Q_{Y} B_\mu + G_X X_\mu \end{pmatrix}}.
\end{equation}
Here, we have defined
\begin{equation}
\label{GXeq}
G_{Xi} = \dfrac{g ' \epsilon Q_{Y_i}}{\cos \theta _W} + g_X Q_{X_i}
\end{equation}
with $Q_{Y_i}$ being the hypercharge of the scalar doublet, which in the 2HDM is taken equal to $+1$ for both scalar doublets, and $Q_{X_i}$ being the charge of the scalar doublet $i$ under $U(1)_X$.
From the part of the Lagrangian responsible for the gauge boson masses, we can extract the relevant terms
\begin{equation}
\begin{split}
\mathcal{L} _{\rm mass} &= \scalemath{0.8}{m_W ^2 W_\mu ^- W ^{+ \mu} + \frac{1}{2} m_{Z} ^2 Z_\mu Z^{\mu} - \Delta ^2 Z_\mu X^\mu + \frac{1}{2} m_X ^2 X_\mu X ^\mu}
\end{split}
\end{equation}
with
\begin{eqnarray}
m_W ^2 = \frac{1}{4} g^2 v ^2 &,& m_{Z} ^2 = \frac{1}{4} g_Z ^2 v ^2
\label{Eq:MZ0}
\end{eqnarray}
and
\begin{equation}
\Delta ^2 = \frac{1}{4} g_Z \left( G_{X1} v_1 ^2 + G_{X2} v_2 ^2 \right)
\label{Eq:Delta2}
\end{equation}
as well as
\begin{equation}
m_X ^2 = \frac{1}{4} \left( v_1 ^2 G_{X1} ^2 + v_2 ^2 G_{X2} ^2 + v_S ^2 g_X ^2 q_X ^2 \right) .
\label{Eq:MX}
\end{equation}
This leads to the gauge boson mass matrix
\begin{equation}
m_{ZX} ^2 = 
\scalemath{0.8}{\frac{1}{8} \begin{pmatrix} g_Z ^2 v ^2 & - g_Z \left( G_{X1} v_1 ^2 + G_{X2} v_2 ^2 \right) \\ - g_Z \left( G_{X1} v_1 ^2 + G_{X2} v_2 ^2 \right)  & v_1 ^2 G_{X1} ^2 + v_2 ^2 G_{X2} ^2 + v_S ^2 g_X ^2 q_X ^2 \end{pmatrix}} \,.
\label{mixinzx}
\end{equation}
The above expression, Eq.\ \eqref{mixinzx}, representing the mixing between the SM $Z$-boson and the new gauge boson $X$, is valid for arbitrary $U(1)_{X}$ charges of singlet and doublet scalars. It is important to notice that, when $Q_{X1}=Q_{X2}$ and there is no singlet contribution, the determinant of the matrix Eq.\ \eqref{mixinzx} is zero.
The matrix in Eq.\ \eqref{mixinzx} is diagonalized through a rotation $O(\xi)$
\begin{equation}
\label{rotacao_zz_fisicos}
\begin{pmatrix} Z \\ Z '  \end{pmatrix} = \begin{pmatrix} \cos \xi & - \sin \xi \\ \sin \xi & \cos \xi \end{pmatrix} \begin{pmatrix} Z \\ X \end{pmatrix}.
\end{equation}
Its mass eigenvalues are
\begin{eqnarray}
\label{autovalores_matriz_zz}
m_{Z} ^2 &=& \frac{1}{2} \left[ m_{Z} ^2 + m_X ^2 - \sqrt{ \left( m_{Z} ^2 - m_X^2 \right) ^2 + 4 \left( \Delta ^2 \right) ^2} \right], \\
m_{Z '} ^2 &=& \frac{1}{2} \left[ m_{Z} ^2 + m_X ^2 + \sqrt{ \left( m_{Z} ^2 - m_X^2 \right) ^2 + 4 \left( \Delta ^2 \right) ^2} \right],
\end{eqnarray}
while the mixing angle $\xi$ is determined by
\begin{equation}
\tan \xi = \frac{ \Delta ^2}{m^2_{Z} - m^2_{X}} .
\label{Eq:xi}
\end{equation}
Since this mixing angle is supposed to be small, as $m_{Z^\prime}^2 \gg m_Z^2$, we can approximate $\tan \xi$ with
\begin{equation}
\sin \xi \simeq \frac{ G_{X1} v_1^2 + G_{X2} v_2^2}{m^2_{Z^\prime}}
\label{Eq:xinew}
\end{equation}
and expand this equation further, substituting the expressions for $G_{Xi}$ and factoring out the $m_Z$ mass, to obtain
\begin{equation}
\sin \xi \simeq \frac{m_Z^2}{ m^2_{Z^{\prime}}}\left(\frac{g_X}{g_Z}(Q_{X1}\cos^2 \beta + Q_{X2}\sin^2 \beta) +\epsilon \tan\theta_W \right).
\label{eqsinxi}
\end{equation}

\bibliography{literature}

\end{document}